\documentstyle{article}
\newskip\humongous \humongous=0pt plus 1000pt minus 1000pt

\newif\ifdtup

	\def\cf{\hbox{\it cf.}{}}

\def\beq{\begin{equation}}
\def\eeq{\end{equation}}

\def\beqn{\begin{eqnarray}}
\def\eeqn{\end{eqnarray}}
\relax

\def\dotx{\dotx{\dot\overline{x}}}

\relax

\def\cf{{\it cf.\ }}

\def\half{\mbox{\small $\frac{1}{2}$}}

\newcommand{\slsh}{\rlap{$\;\!\!\not$}}     % Feynman slash
\long\def\query#1{%
\hskip 0pt{\vadjust{\everypar={}\small\vtop to 0pt{\hbox{}%
\vskip -13pt\rlap{\hbox to 38.5pc{\hfil{\vtop{\hsize=8pc\tolerance=6000%
\hfuzz=.5pc\rightskip=0pt plus 3em\noindent#1}}}}\vss}}}%
}

\def\reflist{\section*{References
\markright{REFERENCES}
}\normalsize\list
        {[\normalsize\arabic{enumi}]\hfill}{\settowidth\labelwidth{[999]}
        \leftmargin\labelwidth
        \advance\leftmargin\labelsep\usecounter{enumi}}}

\relax
\relax
\def\pl#1#2#3{
        {\it Phys.\ Lett.\ }{\bf #1} (19#3) #2}

\def\zp#1#2#3{
        {\it Zeit.\ Phys.\ }{\bf #1} (19#3) #2}

\def\rmp#1#2#3{
        {\it Rev.\ Mod.\ Phys.\ }{\bf #1} (19#3) #2}

\def\pr#1#2#3{
        {\it Phys.\ Rev.\ }{\bf #1} (19#3) #2}
\def\np#1#2#3{
        {\it Nucl.\ Phys.\ }{\bf #1} (19#3) #2}

\def\nc#1#2#3{
        {\it Nuovo Cimento } {\bf #1} (19#3) #2}
\def\JETP#1#2#3{
        {\it Sov.\ Phys.\ JETP} {\bf #1} (19#3) #2}
\def\sj#1#2#3{
        {\it Sov.\ J.\ Nucl.\ Phys.\ } {\bf #1} (19#3) #2}

\relax

\def\half{\mbox{\small $\frac{1}{2}$}}

\def\ason2pi{\frac{\alpha_s}{2 \pi}}
\def\nc{N_C}
\def\as{\alpha_s}

\def\lp{l_+}

\begin{document}
\renewcommand{\floatpagefraction}{0.8}
\begin{titlepage}
\begin{center}
\thispagestyle{empty}
\vskip 1.5cm
\begin{flushright}
         {\vspace*{-0.5in}
         RAL-TR--96-020 \\
         March 1996 \\}
\end{flushright}
\vskip 3 cm
{\Large \bf The spin-dependent two-loop splitting functions\\}
\vskip 2.5 cm
{\large \bf W. Vogelsang}
\vskip 0.3 cm
Rutherford Appleton Laboratory, \\
Chilton, DIDCOT,\\
Oxon OX11 0QX, England. \\
\vskip 3 cm
\today \\
\vskip 3.7 cm
{\bf Abstract}\\
\end{center}
\vskip 0.5 cm
We present a complete description of the calculation of the spin-dependent
next-to-leading order splitting functions. The calculation is performed
in the light-cone gauge. We give results for different prescriptions
for the Dirac matrix $\gamma_5$ in $d=4-2 \epsilon$ dimensions and
provide the link to the results in dimensional reduction. 
\end{titlepage}
\section{Introduction}
It has become standard to perform analyses of unpolarized parton 
distributions at next-to-leading order (NLO) accuracy of QCD. An 
indispensable 
ingredient for such analyses are the two-loop splitting functions 
(or anomalous dimensions) which appear in the NLO $Q^2$-evolution
(GLAP [\ref{AlPa},\ref{GLAP}]) equations. Results (in the 
$\overline{\mbox{MS}}$ scheme)
for these have originally been obtained in [\ref{FLOR}], using
the Operator Product Expansion (OPE) formalism, and in [\ref{CFP},\ref{FP}] 
where the somewhat more efficient method developed in [\ref{EGMPR}] was 
employed\footnote{Note that there is a discrepancy between the results of 
[\ref{FLOR}] and [\ref{FP}] which was resolved in [\ref{HamVN}] in favor of 
the calculation of [\ref{FP}].} which is based on the factorization 
properties of mass singularities and on the use of the axial gauge.
In a recent publication [\ref{EV}] we have presented
a detailed description of the calculation in the latter method 
which had never been fully documented. The main aim of our study was
to elucidate the role played by the light-cone gauge and the 
technical aspects related to its use, and thus to underline the
utility of this gauge for perturbative QCD calculations which had 
been questioned in the past [\ref{Leib},\ref{AT}].

The method of [\ref{EGMPR},\ref{CFP},\ref{FP}] has recently also 
been applied to derive [\ref{Vogelsang}]
the {\em polarized} two-loop splitting functions
needed for the NLO $Q^2$-evolution of the 
spin-dependent parton densities of a longitudinally polarized hadron.
Previous OPE results of [\ref{MVN}] were confirmed. It is the purpose 
of this paper to provide a more detailed and complete description 
of our 'polarized' calculation [\ref{Vogelsang}]. Although many details
of the calculation are the same as for the unpolarized case and can 
therefore be found in [\ref{EV}], there is a new ingredient in the 
polarized case which requires a closer inspection: The Dirac matrix
$\gamma_5$ and the totally antisymmetric Levi-Civita 
tensor $\epsilon_{\mu\nu\rho  
\sigma}$ enter the calculation as projectors onto definite helicity
states of the involved longitudinally polarized quarks and gluons.
When dimensional regularization is used, a prescription for dealing with
these (genuinely {\em four}-dimensional) quantities in $d=4-2 \epsilon$
dimensions has to be adopted
which should be free of algebraic inconsistencies. Several such
'$\gamma_5$ schemes', considered to provide a consistent regularization,
have been suggested in the literature [\ref{HVBM},\ref{KORN},\ref{LV}].
Our calculation [\ref{Vogelsang}] was performed using the original 
definitions for $\gamma_5$ and $\epsilon_{\mu\nu\rho\sigma}$ of
[\ref{HVBM}] (HVBM scheme) which is usually regarded as the most 
reliable prescription. Besides giving a more detailed account of our
previous calculation we will also address the use of other $\gamma_5$
schemes such as [\ref{KORN},\ref{LV}]. Furthermore, we will 
provide the link to the results in dimensional reduction.
\def\Proj{\Delta {\cal P}}
\section{The Calculation}
\subsection{Framework}
An outline of the method of [\ref{EGMPR},\ref{CFP},\ref{FP}] to calculate
(NLO) splitting functions as well as a detailed description of the 
calculation in the unpolarized case have recently been given in
[\ref{EV}]. We will therefore only focus on the new aspects arising in the
polarized case, some of which have already been discussed in our 
previous paper [\ref{Vogelsang}]. 

As usual, all polarized quantities like cross sections etc. will be denoted 
by a '$\Delta$', i.e.,
\beq \label{deltam}
\Delta M \equiv \frac{1}{4} \left( M (++) + M(--) - M(+-) -
M(-+) \right) \; ,
\eeq
where '$+,-$' stand for the helicities of the scattering incoming particles.
The polarized parton distributions $\Delta f$ ($f=q,g$) are defined by 
\beq \label{deltaf}
\Delta f \equiv f^+ - f^- \; ,
\eeq
$f^+$ ($f^-$) denoting the density of parton-type $f$ with positive 
(negative) helicity in a nucleon with positive helicity. Omitting the 
'$\Delta$' in Eqs.~(\ref{deltam},\ref{deltaf}) and taking the sum on the 
right-hand-sides, one recovers the analogous relations for the unpolarized
cross sections and parton distributions.

The general strategy consists of first expanding the squared matrix element 
$\Delta M$ for polarized virtual photon--polarized  quark (gluon) scattering 
into a ladder of two-particle irreducible (2PI) kernels [\ref{EGMPR}]
$C_0$, $K_0$,
\beq \Delta M = \Delta \Bigg[ C_0
(1+K_0+K_0^2+K_0^3+\ldots) \Bigg] \equiv \Delta \Bigg[ \frac{C_0}{1-K_0} 
\Bigg] \; .
\eeq 
We now choose the light-cone gauge by introducing a light-like 
vector $n$ ($n^2=0$) with $n \cdot A=0$. At the same time,
$n$ is used to define the longitudinal direction:
\beq
n \cdot p \equiv pn \neq 0, \;\; n \cdot t=p \cdot t=0 \; ,
\eeq
where $p$ is the momentum of the incoming parton (taken to be massless),
and $t$ is any vector in the transverse plane. In the light-cone gauge 
the 2PI kernels are finite before the integration over the sides of the 
ladder is performed. Collinear singularities therefore appear only when 
integrating over the lines connecting the rungs of the ladder [\ref{EGMPR}]. 
This allows for projecting out the singularities by introducing the 
projector onto polarized physical states, $\Proj$. More precisely, 
$\Proj$ decouples the product, $\Delta (AB)$, of two successive 2PI kernels 
by projecting onto definite helicity states of the particle connecting the 
kernels and by setting this particle on-shell in $A$. Writing down 
explicitly the combinations of the 
helicities of the in- and outgoing partons and of the 
intermediate particle, one immediately obtains from Eq.~(\ref{deltam})
$\Delta ( A \Proj B ) \rightarrow \Delta A \Delta B$, i.e., the 
(decoupled) product of {\em two polarized} kernels. Thus $\Delta M$
can be written in the factorized form 
\beq
\Delta M = \Delta C \Delta \Gamma \;,
\eeq
where (introducing the modified kernel $K=K_0 (1-(1-\Proj )K_0)^{-1}$)
\beqn 
\Delta C&=&\Delta C_0 \frac{1}{1-(1-\Proj )K_0} \; , \label{cexp} \\
\Delta \Gamma &=& \frac{1}{1-\Proj K}  \; , \nonumber \\
&\equiv& 1+\Proj K_0 +\Proj K_0 ( 1-\Proj )K_0 +
(\Proj K_0)(\Proj K_0) +\ldots  \; . \label{gamexp} 
\eeqn
$\Delta C$ is interpreted as the (finite) short-distance cross section,
whereas $\Delta \Gamma$ contains all (and only) mass singularities.
Working in dimensional regularization ($d=4-2 \epsilon$) in the 
$\overline{\mbox{MS}}$ scheme one has 
explicitly:
\beq \label{gam}
\Delta \Gamma_{ij}(x,\as,\frac{1}{\epsilon})=Z_{j} \Bigg[\delta(1-x) 
\delta_{ij}+x \; \mbox{PP} \int \frac{d^dk}{(2 \pi)^d}
\delta(x-\frac{n \cdot k}{pn}) \Delta U_i K \frac{1}{1-\Proj K} 
\Delta L_j\Bigg]  \; ,
\eeq 
where `PP' extracts the pole part of the expression on its right
and $Z_j$ ($j=q(g)$) is the residue of the pole of the full quark (gluon) 
propagator. $k$ is the momentum of the parton leaving the 
uppermost kernel in 
$\Delta \Gamma$; by definition of $n$, $x$ can be interpreted as the 
infinite-momentum frame (IMF) momentum fraction of $p$ carried by $k$. 
The spin-dependent projection operators onto physical states are given by
\beqn 
&&\Delta U_q =-\frac{1}{4 n \cdot k} \gamma_5 \slsh{n},\;\;\;
\Delta L_q = -\slsh{p} \gamma_5 \nonumber \\ 
&&\Delta U_g = i \epsilon^{\mu\nu\rho\sigma} \frac{n_{\rho} k_{\sigma}}
{n \cdot k} , \label{proje} \;\;\; 
\Delta L_g = i \epsilon^{\mu\nu\rho\sigma} \frac{p_{\rho} n_{\sigma}} 
{2pn} \; .
\eeqn
We see that the quantities $\gamma_5$ and $\epsilon^{\mu\nu\rho\sigma}$
appear, for which we will have to define a continuation to $d$ 
dimensions.

Finally, it can be shown [\ref{CFP}] that the coefficient of 
the $1/\epsilon$ pole of $\Delta \Gamma$ corresponds to the GLAP 
[\ref{AlPa},\ref{GLAP}] evolution kernels we are looking for:
\beq
\Delta \Gamma_{qq} (x,\as,\epsilon) = \delta (1-x) - \frac{1}{\epsilon}
\Bigg(\frac{\as}{2\pi} \Delta P_{qq}^{(0)}(x)+\frac{1}{2} 
\left( \frac{\as}{2\pi}
\right)^2 \Delta P_{qq}^{(1)} (x) + \ldots \Bigg) + O \left(
\frac{1}{\epsilon^2} \right) \; ,
\eeq
and analogously for the flavor singlet case. Here we have adopted the 
perturbative expansion of the splitting functions,
\beq \label{expan}
\Delta P_{ij} (x,\as) = \left( \frac{\as}{2\pi} \right) \Delta 
P_{ij}^{(0)} (x) + \left( \frac{\as}{2\pi} \right)^2 
\Delta P_{ij}^{(1)} (x) + \ldots   \; .
\eeq

We conclude this section by collecting all ingredients for a NLO 
study of longitudinally polarized deep-inelastic scattering in terms
of the spin-dependent structure function $g_1 (x,Q^2)$.
There are two different short-distance cross sections\footnote{We do not
distinguish between a quark non-singlet and a quark singlet short-distance
cross section here.}, $\Delta C_q$ and $\Delta C_g$, for scattering off 
incoming polarized quarks and gluons, respectively. They are to be 
calculated according to Eq.~(\ref{cexp}). Thus $g_1$ reads to 
next-to-leading order:
\beqn
g_1 (x,Q^2) &=& \frac{1}{2} \sum_{i=1}^{n_f} e_i^2\; 
\Bigg\{ \Delta q_i (x,Q^2)+\Delta \bar{q}_i (x,Q^2)+ \nonumber \\
&+& \frac{\alpha_s(Q^2)}{2\pi} \left[ \Delta C_q \otimes 
\left( \Delta q_i +\Delta \bar{q}_i \right) +\frac{1}{n_f} 
\Delta C_g \otimes \Delta g\right] (x,Q^2) \Bigg\} \; ,  \label{g1}
\eeqn
where $n_f$ is the number of flavors and $\otimes$ denotes the convolution 
\beq
f \otimes g \equiv \int_0^1 dy \; dz \; f(y) g(z) \; \delta(x-yz)  \; .
\eeq
Defining the sum and the difference of polarized quark and antiquark 
distributions as 
\beq \label{qplus}
\Delta q^\pm_i = \Delta q_i \pm \Delta \bar{q}_i   \; ,
\eeq
one finds the following evolution equations for the non-singlets
$\Delta q_i^-$ and $\Delta q_i^+ - \Delta q_j^+$ (see, e.g., [\ref{EV}]):
\beqn
\frac{d}{d\ln Q^2} (\Delta q_i^+ - \Delta q_j^+) (x,Q^2) &=& 
\Delta P_{qq}^+ (x,\as (Q^2)) \otimes (\Delta q_i^+ - \Delta q_j^+) (x,Q^2)
\; , \label{a3evol} \\
\frac{d}{d\ln Q^2} \Delta q_i^- (x,Q^2) &=& 
\Delta P_{qq}^- (x,\as (Q^2)) \otimes \Delta q_i^- (x,Q^2) \; ,  
\eeqn
where
\beq   \label{pm}
\Delta P_{qq}^{\pm} \equiv \Delta P^V_{qq} \pm \Delta P^V_{q\bar{q}}   \; ,
\eeq
with $\Delta P_{q\bar{q}}^V$ starting to be different from zero beyond 
the leading order.
Introducing the polarized quark singlet $\Delta \Sigma \equiv \sum_i 
(\Delta q_i+\Delta \bar{q}_i)$ one has in the singlet sector:
\begin{eqnarray} \label{AP}
\frac{d }{d \ln Q^2}   
\left( \begin{array}{c} \Delta \Sigma (x,Q^2) \\ \Delta g (x,Q^2)  
\end{array} \right)
= \left( \begin{array}{cc} \Delta P_{qq}(x,\as (Q^2)) &  
\Delta P_{qg}(x,\as (Q^2))\\  
\Delta P_{gq}(x,\as (Q^2) ) &  \Delta P_{gg}(x,\as (Q^2) )\\  
\end{array}\right)  \otimes
\left(  \begin{array}{c}
\Delta \Sigma(x,Q^2) \\ 
\Delta g(x,Q^2)
\end{array} \right) \; .
\end{eqnarray}
The $qq$ entry in the singlet matrix of splitting functions is given 
by\footnote{As compared to [\ref{EV}] we include a factor $2 n_f$ 
in the definition of $\Delta P_{qq}^S$.}
\beq  \label{qqs}
\Delta P_{qq}=\Delta P_{qq}^+ + \Delta P^S_{qq} \; .
\eeq
So, at NLO, we will have to derive the splitting functions 
$\Delta P_{qq}^{\pm ,(1)}$, $\Delta P_{qq}^{S, (1)}$, 
and those involving gluons.
\subsection{NLO graphs}
\begin{figure}[htb]
\vspace{6.5cm}
\includegraphics{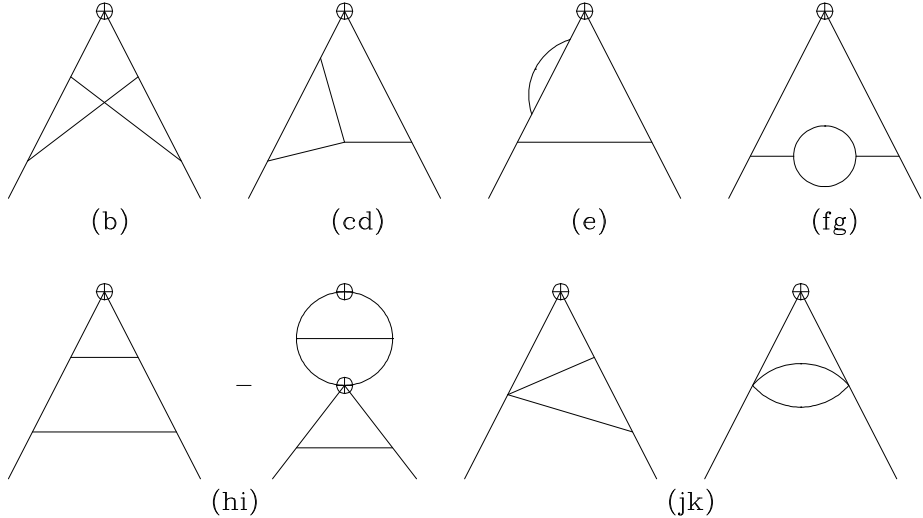}
\vspace*{-0.3cm}
\caption{Basic topologies of the NLO diagrams}
\vspace*{-0.2cm}
\label{topol}
\end{figure}

According to Eqs.~(\ref{gamexp},\ref{gam}), $\Delta \Gamma$ 
is given to NLO by
\beq 
\Delta \Gamma = Z_j \left( 1+\Proj K_0 +\Proj (K_0^2) - \Proj (K_0 
\Proj K_0) \right) . \label{gamma} 
\eeq
The basic topologies of all 2PI diagrams which occur in NLO are shown 
in Fig.~\ref{topol}, where the notation of the topologies is as in 
[\ref{EV}]. 
Explicit examples of graphs contributing to the various splitting functions 
can be found in [\ref{CFP},\ref{Vogelsang},\ref{EV}]. 
Topologies (hi) correspond to the terms 
$\Proj (K_0^2)-\Proj (K_0 \Proj K_0)$ 
in Eq.~(\ref{gamma}), all other topologies belong to $\Proj K_0$.
As is obvious, topologies (cd) and (fg) possess a real and a virtual 
cut. Fig.~\ref{topol} does not display the genuine two-loop graphs which 
determine $Z_j$ and thus the endpoint ($\delta (1-x)$) contributions 
to the diagonal splitting functions.
\vspace*{0.5cm}
\subsection{Use of the light-cone gauge}
As noted in the previous section, the light-cone gauge plays 
a crucial role in the calculation. In this gauge the gluon 
propagator takes the form 
\beq
{\cal D}^{\mu\nu} (l) = \frac{i}{l^2} \Bigg( -g^{\mu \nu}+
\frac{n^\mu l^\nu+n^\nu l^\mu}{n\cdot l}  \Bigg) \; ,
\eeq
where $l$ is the gluon's momentum. As is well-known, the light-cone 
gauge propagator $1/(n\cdot l)$ can give rise to additional divergencies
in loop and phase space integrals. We follow [\ref{CFP},\ref{FP},\ref{EV}]
to use the principal value (PV) prescription to regulate such poles:
\beq \label{PPprescription}
\frac{1}{n\cdot l} \rightarrow \frac{1}{2} \Bigg(
\frac{1}{n\cdot l+ i \delta (pn)} + 
\frac{1}{n\cdot l- i \delta (pn)} \Bigg)
=\frac{n \cdot l}{(n\cdot l)^2 + \delta^2 (pn)^2} \; .
\eeq 
All singularities related to the gauge propagator can then be cast into
the basic integrals [\ref{CFP},\ref{EV}]
\beq
I_i = \int_0^1 du \frac{u \ln^i u}{u^2+\delta^2} \;\;\; (i=0,1) \; .
\eeq
The  PV prescription leads to the feature that the 
renormalization constants depend on $I_0$ and the longitudinal
momentum fractions $x$ [\ref{CFP},\ref{EV}]. The complete set of the 
renormalization constants, as well as most other technical 
ingredients we need for our 
calculation, like phase space integrals and scalar virtual integrals
with and without a gauge propagator, have been worked out and collected 
in [\ref{EV}].  
\subsection{Treatment of $\gamma_5$ and $\epsilon_{\mu\nu\rho\sigma}$}
As seen from Eq.~(\ref{proje}), the crucial difference with respect to 
the unpolarized calculation is the presence of $\gamma_5$ and 
$\epsilon_{\mu\nu\rho\sigma}$ which act as projectors. Definitions 
for these quantities in $d=4-2 \epsilon$ dimensions (or at least rules for 
handling them) have to be given. 

It is well known that a 'naive' fully anticommuting $d$-dimensional
$\gamma_5$ can easily lead to algebraic inconsistencies when located 
in a trace with six or more other Dirac matrices [\ref{CFH}], and 
thus results obtained this way cannot really be trusted. 
The problem can be avoided in essentially two ways: 
\begin{itemize}
\item One can {\em define} $\gamma_5$ in $d$ dimensions by maintaining its 
{\em four}-dimensional definition 
\beq  \label{gamma5}
\gamma_5 \equiv \frac{i}{4!} \epsilon^{\mu\nu\rho\sigma} \gamma_{\mu}
\gamma_{\nu}\gamma_{\rho}\gamma_{\sigma} \; ,
\eeq       
where the $\epsilon$-tensor is regarded as a genuinely four-dimensional
object, i.e., its components vanish in all unphysical dimensions. This 
definition of $\gamma_5$ is the original one of [\ref{HVBM}] (HVBM
scheme). Splitting the $d$-dimensional metric tensor into its
four- and $(d-4)$-dimensional components, 
\beq
g_{\mu\nu} = \hat{\hat{g}}_{\mu\nu} + \hat{g}_{\mu\nu} \;\;\;
(\mbox{where} \; \hat{\hat{g}}_{\mu}^{\mu}=4, \;
\hat{g}_{\mu}^{\mu}= d-4) \; , 
\eeq
and defining $\hat{\gamma}_{\mu} \equiv \hat{g}_{\mu\nu} \gamma^{\nu}$ 
etc., one finds from (\ref{gamma5}) that
\beqn
\left\{ \gamma^{\mu},\gamma_5 \right\} &=& 0  \;\;\; \mbox{for} 
\; \mu=0,1,2,3 \; ,  \nonumber    \\
\left[ \gamma^{\mu},\gamma_5 \right] &=& 0  \;\;\; \mbox{otherwise}  \; .
\label{g5anti}
\eeqn
\item One can maintain the full anticommutativity of $\gamma_5$ with all
other Dirac matrices but abandon cyclicity of the trace operation 
[\ref{KORN}]. In this case a fixed 'reading point' has to be defined from 
which all Dirac traces corresponding to a given process have to be started.
As in the HVBM scheme, the $\epsilon$-tensor is defined as an entirely
four-dimensional object [\ref{KORN}].
\end{itemize}
Our main calculation\footnote{We use the 
program {\sc Tracer} of [\ref{TRAC}] for calculating the Dirac traces and 
performing contractions.} is done in the HVBM scheme since its
consistency is well-established. We will also discuss the results 
obtained in the scheme of [\ref{KORN}] which was used in the OPE 
calculation of the $\Delta P_{ij}^{(1)}$ of [\ref{MVN}].
Due to the fact that the Levi-Civita tensor is assumed to be 
four-dimensional (and because of (\ref{g5anti}) in the HVBM scheme), 
the matrix element of a graph will for both prescriptions 
not only depend
on the usual '$d$-dimensional' scalar products of two momenta, like 
$l \cdot k \equiv g_{\mu\nu} l^{\mu} k^{\nu}$ etc., 
but also on '$(d-4)$-dimensional' ones, denoted by 
$\hat{l}\cdot \hat{k} \equiv \hat{g}_{\mu\nu} l^{\mu} k^{\nu}$, 
$\hat{k}^2$ etc. Special care has to be taken 
to take into account such $(d-4)$-dimensional terms in loop and phase 
space integrals. Details are given in the appendices.

We mention that in [\ref{LV},\ref{NZ}] another prescription 
was suggested to handle traces with one $\gamma_5$. One eliminates 
$\gamma_5$ via the relation
\beq \label{larg5}
\gamma_{\mu} \gamma_5 = \frac{i}{3!} \epsilon_{\mu\nu\rho\sigma}
\gamma^{\nu} \gamma^{\rho} \gamma^{\sigma} \; .
\eeq
The remaining trace is then perfectly well-defined. One always ends up with
the product of two Levi-Civita tensors which can be written as a 
determinant of a matrix, the elements of which are metric tensors.
Following [\ref{LV},\ref{NZ}], one then performs the contractions of these
metric tensors in $d$ dimensions\footnote{Here one has to multiply
the gluonic projection operator $\Delta L_g$ in (\ref{proje}) by a 
normalization factor \linebreak[4] $2/((d-2)(d-3))$.}. 
The result should then coincide with that in the HVBM scheme. The scheme
is attractive because it avoids any $(d-4)$-dimensional scalar
products, but it leads to longer trace operations because of the
substitution (\ref{larg5}). We will briefly return to this prescription when
presenting our results.

We finally note that a variant of dimensional regularization, dimensional
{\em reduction} [\ref{siegel}], has been widely discussed and used in
the past years, the main reason being its applicability to supersymmetry.
The scheme essentially consists of performing the Dirac-algebra in 
{\em four} dimensions, which makes the treatment of $\gamma_5$ 
straightforward, and of continuing only momenta to $d$ ($d<4$) dimensions. 
In order to match the ultraviolet (UV) sectors of dimensional 
regularization and dimensional reduction, specific counterterms have 
to be introduced [\ref{SSK},\ref{KT}] in the latter. Once this is 
done, there is a straightforward and universal way to deal with differences 
arising from mass singularities [\ref{KT},\ref{KST},\ref{KAMAL}]. We will
exploit the results of [\ref{KT},\ref{KST},\ref{KAMAL}] in order to 
translate our results to the form they take in dimensional reduction.
\section{Results}
The full one-loop results are included for completeness 
[\ref{AlPa},\ref{AhRo}]:
\beqn
\label{Pqq}
\Delta P_{qq}^{(0)}(x)&=&C_F \Big\{\frac{2}{[1-x]_+}-1-x 
+\frac{3}{2}\delta(1-x)
\Big\} \\
\label{Pqg}
\Delta P_{qg}^{(0)}(x)&=&2 T_f\left\{2x-1 \right\}  \\
\label{Pgq}
\Delta P_{gq}^{(0)}(x)&=&C_F \left\{ 2-x \right\} \\
\label{Pgg}
\Delta P_{gg}^{(0)}(x)&=&2 \nc 
\Big\{\frac{1}{[1-x]_+}-2 x+1 \Big\}
+\frac{\beta_0}{2} \delta(1-x)  \; ,
\eeqn
where \beq
C_F=\frac{4}{3},\;
\nc=3,\;
T_f=T_R n_f = \half n_f, \; \beta_0 = \frac{11}{3} \nc -\frac{4}{3} T_f \; . 
\eeq
\subsection{Results in the HVBM scheme}
The graph-by-graph results (for those splitting functions that involve more 
than just one topology) in the HVBM scheme are given in columns (b)-(jk) 
of Tables 1-4. For the non-singlet
case, Table 1, we only present the differences between our graph-by-graph
results for the
polarized $\Delta P_{qq}^{V,(1)}$ and the corresponding contributions to
the unpolarized $P_{qq}^{V,(1)}$ as listed in Table 1 of
[\ref{CFP}]. The reason for this is that for an
anticommuting $\gamma_5$ the $qq$ diagrams would trivially yield
$\Delta P_{qq}^{V,(1)} = P_{qq}^{V,(1)}$ because of the fact that 
there are always two $\gamma_5$ in the same trace which could be 
anticommuted towards each other and eliminated using $\gamma_5^2=1$.
However, as discussed in section 2.4, $\gamma_5$ does {\em not} fully
anticommute in the HVBM scheme. So it is interesting to see what happens
to the difference $\Delta P_{qq}^{V,(1)} - P_{qq}^{V,(1)}$ when this
prescription is used. To examine this question further we have also
distinguished in Table 1 the contributions coming from the 
$(d-4)$-dimensional 
scalar products, like $\hat{k}^2$ etc. (see section 2.4 and the appendices),
from all other contributions\footnote{In all other tables the contributions
from the $(d-4)$-dimensional scalar products are already included.}.

Let us concentrate on the columns 'Sum' in each table, 
which give the sums of the graph-by-graph results (b)-(jk). The first thing 
to observe is that all contributions from the integrals $I_0$, $I_1$,
arising from the light-cone gauge propagator, cancel in 'Sum', as
they must. We furthermore see from Table 1 that
$\Delta P_{qq}^{V,(1)} - P_{qq}^{V,(1)}$
is indeed non-vanishing in the terms multiplied by $C_F N_C$ and
$C_F T_f$. To study the implications of this let us first note that 
the endpoint contributions ($\sim \delta(1-x)$) to the diagonal splitting 
functions are necessarily the same as in the unpolarized case 
(where they were derived in [\ref{Wada},\ref{EV}]), since they 
are determined by $Z_j$ in Eq.~(\ref{gam}). We can then express
the results corresponding to the columns 'Sum' in terms of the 
unpolarized NLO non-singlet splitting functions $P_{qq}^{\pm,(1)}$
of [\ref{CFP}] and the recent polarized OPE results $\Delta 
\tilde{P}_{ij}^{(1)}$ of [\ref{MVN}]\footnote{In our normalization 
given by Eq.~(\ref{expan},\ref{g1}) the results for 
the NLO splitting functions of [\ref{MVN}] have to be divided by 8
and those for the NLO short-distance cross sections $\Delta C_q$, 
$\Delta C_g$ by $\as /2\pi$.}:
\begin{eqnarray}
\Delta P_{qq}^{\pm ,(1)} (x) &=& P_{qq}^{\mp ,(1)} (x) - 2 \beta_0 
C_F (1-x) \:\:\: , \label{inter} \\
\Delta P_{qq}^{S,(1)} (x) &=& \Delta \tilde{P}_{qq}^{S,(1)}
(x) \:\:\: ,  \\
\Delta P_{qg}^{(1)} (x) &=& \Delta \tilde{P}_{qg}^{(1)}
(x) + 4 C_F (1-x) \otimes \Delta P_{qg}^{(0)} (x) \:\:\: ,  \\
\Delta P_{gq}^{(1)} (x) &=& \Delta \tilde{P}_{gq}^{(1)}
(x) - 4 C_F (1-x) \otimes \Delta P_{gq}^{(0)} (x) \:\:\: , \\
\Delta P_{gg}^{(1)} (x) &=& \Delta \tilde{P}_{gg}^{(1)} (x)
\:\:\: , \\
\Delta C_q (x) &=& \Delta \tilde{C}_q (x)
-4 C_F (1-x) \:\:\: , \label{cq} \\
\Delta C_g (x) &=& \Delta \tilde{C}_g (x) \:\:\: ,
\label{cg}
\end{eqnarray}
where we have also included the results for the short-distance 
cross sections $\Delta C_q$, $\Delta C_g$ which are to be calculated 
according to Eq.~(\ref{cexp}). Obviously, the term $-2 \beta_0 C_F
(1-x)$ in (\ref{inter}) is entirely due to the fact the HVBM $\gamma_5$
does not fully anticommute. As was already discussed in 
[\ref{grsv},\ref{ms}] and indicated in Eq.~(\ref{inter}), the '$+$'
and '$-$' combinations of the NS splitting functions as 
defined in (\ref{pm}) interchange their role when going from the 
unpolarized to the polarized case,
equivalent to $\Delta P_{q\bar{q}}^{V,(1)}=-P_{q\bar{q}}^{V,(1)}$.
The latter relation is a consequence of the projection 
operator for antiquarks [\ref{CFP}], $+\gamma_5 (-\slsh{n})/4 n\cdot k$, 
in this case. Again it is trivial for a fully anticommuting $\gamma_5$, 
but it turns out that it is also respected by the HVBM $\gamma_5$.
Eqs.~(\ref{a3evol},\ref{pm},\ref{inter}) therefore imply that
the combination $\Delta P_{qq}^{+,(1)}=P_{qq}^{-,(1)} - 2\beta_0 C_F (1-x)$ 
would govern the $Q^2$-evolution of, e.g., the polarized NS quark 
combination 
$$\Delta A_3 (x,Q^2) = \left( \Delta u^+ - \Delta d^+ \right) (x,Q^2)
\:\:\: .$$
Since the first moment (i.e., the $x$-integral) of the latter corresponds
to the nucleon matrix element of the NS axial vector current
$\bar{q} \gamma^{\mu} \gamma_5 \lambda_3 q$ which is conserved, 
it has to be $Q^2$-independent [\ref{kod}]. Keeping in mind that 
the integral of the unpolarized $P_{qq}^{-,(1)}$
vanishes already due to fermion number conservation 
[\ref{CFP}], it becomes obvious that the additional term $-2 \beta_0
C_F (1-x)$ in (\ref{inter}) spoils the $Q^2$-independence of the first 
moment of $\Delta A_3 (x,Q^2)$.
It is therefore necessary to perform a factorization scheme transformation
to the results in (\ref{inter}-\ref{cg}) in order to remove this additional
term. Such a scheme transformation is always allowed since neither
the $\Delta P_{ij}^{(1)}$ nor the $\Delta C_i$ are physical quantities. 
Thus one can shift terms between them in a well-defined way 
without changing a physical quantity like $g_1$, hereby just redefining 
the polarized 
NLO parton distributions. Even though it is a priori only the term 
$-2 \beta_0 C_F (1-x)$ in (\ref{inter}) we want to remove, the 
transformation will also affect the singlet sector since, according
to Eq.~(\ref{qqs}), $\Delta P_{qq}^{+,(1)}$ also occurs in the 
singlet evolution matrix. The scheme transformation reads in general (see,
e.g., [\ref{GR},\ref{MVN}]):
\beqn
\Delta P_{qq}^{\pm,(1)} &\longrightarrow& \Delta P_{qq}^{\pm,(1)}
-2 \beta_0 \Delta z_{qq}  \; , \nonumber \\
\Delta P_{qq}^{(1)} &\longrightarrow& \Delta P_{qq}^{(1)}
-2 \beta_0 \Delta z_{qq} + 4 \Delta z_{qg} \otimes \Delta P_{gq}^{(0)} -
4 \Delta z_{gq} \otimes \Delta P_{qg}^{(0)} \; , \nonumber \\
\Delta P_{qg}^{(1)} &\longrightarrow& \Delta P_{qg}^{(1)}
-2 \beta_0 \Delta z_{qg} + 4 \Delta z_{qg} \otimes \left( 
\Delta P_{gg}^{(0)} - \Delta P_{qq}^{(0)} \right) 
+ 4 \Delta P_{qg}^{(0)} \otimes \left( \Delta z_{qq} - \Delta z_{gg}
\right) \; , \nonumber \\
\Delta P_{gq}^{(1)} &\longrightarrow& \Delta P_{gq}^{(1)}
-2 \beta_0 \Delta z_{gq} + 4 \Delta z_{gq} \otimes \left( 
\Delta P_{qq}^{(0)} - \Delta P_{gg}^{(0)} \right) 
+ 4 \Delta P_{gq}^{(0)} \otimes \left( \Delta z_{gg} - \Delta z_{qq}
\right) \; , \nonumber \\
\Delta P_{gg}^{(1)} &\longrightarrow& \Delta P_{gg}^{(1)}
-2 \beta_0 \Delta z_{gg} - 4 \Delta z_{qg} \otimes \Delta P_{gq}^{(0)} +
4 \Delta z_{gq} \otimes \Delta P_{qg}^{(0)} \; , \nonumber \\
\Delta C_q &\longrightarrow& \Delta C_q - 4 \Delta z_{qq} \; , \nonumber \\
\Delta C_g &\longrightarrow& \Delta C_g - 4 \Delta z_{qg} \; , 
\label{strafo}
\eeqn
where the $\Delta z_{ij}$ generate the transformation. For the case
at hand we only need a non-vanishing $\Delta z_{qq}$, 
\beq  \label{zqq}
\Delta z_{qq} (x) = -C_F (1-x) \; .
\eeq
The changes to our result caused by inserting $\Delta z_{qq}$ into
(\ref{strafo}) are given in the columns ''$\gamma_5$'' in Tables 1-3, to be
added to columns 'Sum' to obtain the final answer. As one can see from 
Table 1, the difference $\Delta P_{qq}^{V,(1)} - P_{qq}^{V,(1)}$
finally becomes zero, which is a result of cancellations between terms 
from $(d-4)$-dimensional scalar products and other terms. Furthermore,
it turns out that 
the transformation (\ref{strafo},\ref{zqq}) not only removes the 
term $-2 \beta_0 C_F (1-x)$ from Eq.~(\ref{inter}), but eliminates 
{\em all} extra $(1-x)$-terms on the r.h.s. of (\ref{inter}-\ref{cg}), 
leaving $\Delta P_{qq}^{S,(1)}$, $\Delta P_{gg}^{(1)}$ and $\Delta C_g$
unchanged. Thus our final results are in complete agreement with those 
of [\ref{MVN}]. We finally note that the presence of the $(1-x)$-terms 
in our original HVBM scheme result (\ref{inter}-\ref{cg}) can be traced
back to the fact that in this scheme the polarized {\em LO} splitting 
function in $d=4-2 \epsilon$ dimensions, $\Delta P_{qq}^{(0),d=4-2 
\epsilon}$, is no longer equal to its unpolarized counterpart, i.e., 
violates helicity conservation:
\beq \label{pqqdim}
\Delta P_{qq}^{(0),d=4-2 \epsilon} (x)-P_{qq}^{(0),d=4-2 \epsilon} (x) =
4 \epsilon (1-x) \; .
\eeq
The additional term $-4 C_F (1-x)$ in the HVBM-scheme result for $\Delta
C_q$ in (\ref{cq}) was already identified in [\ref{Alex},\ref{ms}]. After
its removal by the transformation (\ref{strafo},\ref{zqq}), the integral 
over $\Delta C_q$ takes the value $-3 C_F/2$, giving rise to the
correct NLO correction ($1-\as/\pi$) to, e.g., the Bj\o rken sum rule.

Our complete final results can now be collected from columns 'final'
(or 'Sum' if there has been no change due to (\ref{strafo},\ref{zqq})) in 
Tables~1-4. To write them down we introduce
\begin{eqnarray}
\delta p_{qg} (x) &\equiv& 2 x-1  \:\:\: , \nonumber \\
\delta p_{gq} (x) &\equiv& 2 - x  \:\:\: , \nonumber \\
\delta p_{gg} (x) &\equiv& \frac{1}{[1-x]_+} - 2 x + 1  \label{pgg}  \:\:\: .
\end{eqnarray}
We then have
\begin{eqnarray}
\Delta P_{qq}^{\pm,(1)} &=& P_{qq}^{\mp,(1)}  \:\:\: , \label{p1ns} \\
\Delta P_{qq}^{S,(1)} (x) &=& 2 C_F T_f \Bigg[ \left( 1 - x \right)
-\left( 1 - 3 x \right) \ln x  - \left( 1 + x \right) 
\ln^2 x \Bigg]  \:\:\: , \label{p1s} \\
\Delta P_{qg}^{(1)} (x) &=& C_F T_f \Bigg[ -22 + 27 x - 9 \ln x + 
     8 \left( 1 - x \right) \ln (1-x) \nonumber \\
&& + 
     \delta p_{qg}(x) \left( 2 \ln^2 (1-x) - 
      4 \ln (1-x) \ln x + \ln^2 x - \frac{2}{3} \pi^2 \right) 
     \Bigg] \nonumber \\
&&+ N_C T_f \Bigg[ 2 \left( 12 - 11 x \right)  - 
     8 \left( 1 - x \right) \ln (1-x) + 2 \left( 1 + 8 x \right) \ln x 
     \nonumber \\ 
&& \left. - 2 \left( \ln^2 (1-x) - \frac{\pi^2}{6} \right) \delta p_{qg}(x) 
- \left( 2 S_2 (x) - 3 \ln^2 x \right)  \delta p_{qg}(-x) \Bigg] \right. 
\:\:\: , \label{p1qg} \\
\Delta P_{gq}^{(1)} (x) &=& C_F T_f \left[ -{{4}\over 9} (x+4)
- \frac{4}{3} \delta p_{gq}(x) \ln (1-x) \right] \nonumber 
\\ &&+ C_F^2 
   \left[ - \frac{1}{2} -
   \frac{1}{2} \left( 4 - x \right) \ln x - \delta p_{gq}(-x) \ln (1-x)
\right. \nonumber \\
&& \left. + \left( - 4 - \ln^2 (1-x) + \frac{1}{2} \ln^2 x \right)  
    \delta p_{gq}(x) \right] \nonumber \\
&& + C_F N_C \left[ \left( 4 - 13 x \right) \ln x + 
   \frac{1}{3} \left( 10 + x \right) \ln (1-x) + 
  \frac{1}{9} \left( 41 + 35 x \right) \right. \nonumber \\
&& + \frac{1}{2} \left( -2 S_2 (x) + 3 \ln^2 x \right)  
     \delta p_{gq}(-x) \nonumber \\
&&+ \left. \left( \ln^2 (1-x) - 2 \ln (1-x) \ln x - \frac{\pi^2}{6} \right)
  \delta p_{gq}(x) \right]  \label{p1gq} \\
\Delta P_{gg}^{(1)} (x) &=& 
- N_C T_f \left[ 4 \left( 1 - x \right)  + \frac{4}{3} 
     \left( 1 + x \right) \ln x + \frac{20}{9} \delta p_{gg}(x)
     +\frac{4}{3} \delta (1-x) \right] \nonumber \\
&& - C_F T_f 
  \Bigg[ 10 \left( 1 - x \right)  + 2 \left( 5 - x \right) \ln x  + 
     2 \left( 1 + x \right) \ln^2 x  
     +\delta (1-x) \Bigg]  \nonumber \\
&& + N_C^2 \Bigg[ \frac{1}{3} \left( 29 - 67 x \right) \ln x -
     \frac{19}{2} \left( 1 - x \right) + 
     4 \left( 1 + x \right) \ln^2 x -2 S_2 (x) \delta p_{gg}(-x) 
\nonumber \\
&& + \left( {{67}\over 9} - 4 \ln (1-x) \ln x + 
        \ln^2 x - \frac{\pi^2}{3} \right) \delta p_{gg}(x) 
 +\left( 3 \zeta(3) +\frac{8}{3} \right) \delta (1-x) \Bigg] \; , 
\nonumber \\ 
\label{p1gg} \\  
\Delta C_q(x) &=& C_F \Bigg[(1+x^2) \left[\frac{\ln (1-x)}{1-x}
\right]_{\!\!+}
-\frac{3}{2} \frac{1}{[1-x]_+} -\frac{1+x^2}{1-x} \ln x +
\nonumber \\
& & \hspace*{0.75cm} 
+\, 2 + x - \left(\frac{9}{2}+\frac{\pi^2}{3}\right) \delta (1-x)
\Bigg]  \label{cqend} \:\:\: ,  \\
\Delta C_g(x) &=& 2 T_f \left[ (2x-1) \left(\ln \frac{1-x}{x}-1\right)+
2(1-x)\right] \:\:\: \label{cgend} ,
\end{eqnarray}
where the unpolarized NS pieces $P_{qq}^{\mp, (1)}$ can be found 
in [\ref{CFP},\ref{EV}], and
\beq
S_2(x)= \int_{\frac{x}{1+x}}^{\frac{1}{1+x}} \frac{dz}{z} 
\ln \big(\frac{1-z}{z}\big)  \; .
\eeq                        
For relating the above results to those of [\ref{MVN}] the relation 
\beq 
S_2 (x) = -2 {\rm Li}_2 (-x)-2 \ln x \ln (1+x)+\frac{1}{2} \ln^2 x-
\frac{\pi^2}{6} 
\eeq
is needed, where ${\rm Li}_2 (x)$ is the Dilogarithm [\ref{dd}]. 
As explained above, the $\delta (1-x)$-endpoint contributions 
could be taken from the unpolarized case [\ref{Wada},\ref{EV}]; 
$\zeta (3)\approx 1.202057$.
The $+$-prescription in (\ref{pgg},\ref{cqend}) is defined in the usual way,
\beq
\int_0^1 dz f(z) \left[ g(z) \right]_+ 
\equiv \int_0^1 dz \left( f(z)-f(1) \right) g(z) \:\:\: . 
\eeq
It is obviously only needed if the function multiplying it is non-vanishing
at $x=1$. 

For completeness we finally list the first moments\footnote{Compact 
expressions for all Mellin-moments of the NLO splitting functions, 
$\int_0^1 dx x^{n-1} \Delta P_{ij}^{(1)} (x)$, and their analytic 
continuations, can be found in [\ref{grsv}].} ($x$-integrals) of the results 
in Eqs.~(\ref{p1s}-\ref{cgend}):
\beqn
&&\int_0^1 \Delta P_{qq}^{S,(1)} (x) dx = -3 C_F T_f  \;\; , \;\;\;  
\int_0^1 \Delta P_{gq}^{(1)} (x) dx = -\frac{9}{4} C_F^2 + 
\frac{71}{12} N_C C_F - \frac{1}{3} C_F T_f \; , \nonumber \\ 
&&\int_0^1 \Delta P_{qg}^{(1)} (x) dx = 0  \;\; , \;\;\; 
\int_0^1 \Delta P_{gg}^{(1)} (x) dx = \frac{17}{6} N_C^2 - C_F T_f
-\frac{5}{3} N_C T_f \equiv \frac{\beta_1}{4} \; , \nonumber \\ 
&&\int_0^1 \Delta C_q (x) dx = -\frac{3}{2} C_F \;\; , \;\;\; 
\int_0^1 \Delta C_g (x) dx = 0 \; .
\eeqn
\subsection{Results for other $\gamma_5$ prescriptions}
\begin{figure}[htb]
\vspace{5cm}
\includegraphics{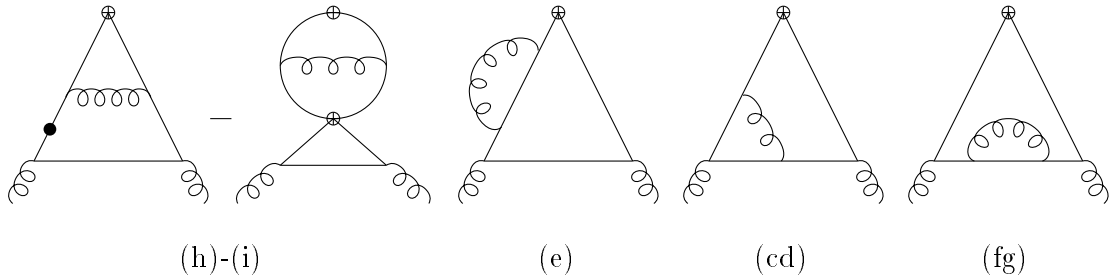}
\caption{Graphs for the $C_F T_f$ part of $\Delta P_{qg}^{(1)}$}
\label{topolqg}
\end{figure}

We now discuss the results we obtain when using the prescription 
of [\ref{KORN}] with an anticommuting $\gamma_5$ and a non-cyclic trace.
As mentioned earlier,
the property $\{ \gamma^{\mu},\gamma_5 \}=0$ will automatically 
yield $\Delta P_{qq}^{\pm,(1)} = P_{qq}^{\mp,(1)}$ instead of (\ref{inter}).
Thus it immediately follows that {\em either} all other $(1-x)$-terms in
(\ref{inter}-\ref{cg}) are absent as well, {\em or} that the full result 
will be 
genuinely different from the one in Eqs.~(\ref{p1ns}-\ref{cgend}), i.e., not
transformable into (\ref{p1ns}-\ref{cgend}) by a factorization 
scheme transformation (\ref{strafo}). To study this question, we have to
calculate the other NLO splitting functions and the short-distance 
cross sections. Since the prescription for the Levi-Civita tensor 
is the same as in the HVBM scheme, it is obvious that $\Delta P_{gg}^{(1)}$ 
and $\Delta C_g$ will be the same in both schemes.
The remaining quantities are the interesting ones since they involve one or
two traces with one $\gamma_5$. In the prescription of [\ref{KORN}], one has 
to define a 'reading point' for these, at which the trace is to be started. 
Let us discuss our choice for the case of the $C_F T_f$ part of 
$\Delta P_{qg}^{(1)}$, in which the incoming particle is a gluon and 
the outgoing one a quark with the projector $-\gamma_5 \slsh{n}/4 n \cdot k$   
being acted upon. Only topologies (cd),(e),(fg) and (hi) contribute; for the 
sake of clarity we show the graphs explicitly in Fig.~\ref{topolqg}. 
It seems reasonable to choose the projector as the reading point, marked by
the 'crosses' $\oplus$ in Fig.~\ref{topolqg}. 
Doing so for graphs (cd),(e),(fg), which in the 
language of Eq.~(\ref{gamma}) represent $\Proj K_0$, 
one obtains for each graph exactly the same answer as when using the
HVBM $\gamma_5$. However, for topology (hi) one has to take care:
For the subtraction graph (i), which represents $-\Proj (K_0 \Proj K_0)$,
the projection operator $\Proj$ has by definition effectively cut 
the Dirac-trace of (h) ($=\Proj(K_0^2)$) into two traces, 
hereby inserting an additional $\gamma_5$ in each trace according to 
$\Delta L_q$, $\Delta U_q$ in (\ref{proje}). Thus (i) essentially becomes 
the convolution of $\Delta P_{qq}^{(0)}$ for the upper part of the 
diagram with $\Delta P_{qg}^{(0)}$ for the lower. Since the trace for the 
upper part then contains 
two $\gamma_5$ which eliminate each other, the $\gamma_5$-problem has 
been shifted from the upper 'cross' to the lower one. This suggests
that the trace for graph (h) should be read from the black dot rather 
than from the 'cross'. Indeed, when doing this, it turns out that 
both pieces (h),(i) 
yield differences with respect to the HVBM results, which
on aggregate are exactly the same as the entries in the column 
''$\gamma_5$'' in Table 2. This means that we end up with the
result in (\ref{p1qg}) for the $C_F N_f$ part of $\Delta P_{qg}^{(1)}$,
without {\em any} extra terms. The same thing happens for the case of the 
$C_F^2$ part of $\Delta P_{gq}^{(1)}$, where it is resonable to choose 
the $-\slsh{p} \gamma_5$ of the incoming polarized quark as the 
reading point for the graphs for $\Proj K_0$, in which case each of 
these graphs individually gives the same result as in the HVBM scheme.
This time, the answer for subtraction graph (i) is also the same as 
in the HVBM scheme, but when the reading point for graph (h) 
($\Proj (K_0^2)$) is again chosen as the quark propagator 
on the side of the ladder between the rungs, it alone 
generates all differences in column ''$\gamma_5$'' of Table~3, such that the 
$C_F^2$ term in (\ref{p1gq}) is reproduced. Next, we have to calculate 
the $N_C T_f$ part of $\Delta P_{qg}^{(1)}$ and the $C_F N_C$ part 
of $\Delta P_{gq}^{(1)}$, choosing again $-\gamma_5 \slsh{n}/4 n \cdot k$
and $-\slsh{p} \gamma_5$ as the reading points, respectively. Here, 
no extra care has to be taken since for both the sides of the ladder between
the rungs in graph (h) are gluons. For each graph the result exactly agrees
with the corresponding one in the HVBM scheme. Finally, we have 
calculated $\Delta P_{qq}^{S,(1)}$, again choosing the projectors as 
the reading points for its two traces and reproducing once again the result 
of the calculation in the HVBM scheme. Also, $\Delta C_q$ in (\ref{cqend})
was obtained in [\ref{ms}] with the photon vertex as reading point.

Thus, to summarize our results for the calculation in the $\gamma_5$-scheme
of [\ref{KORN}], we have indeed reproduced all polarized NLO quantities
as given in their final form in Eqs.~(\ref{p1ns}-\ref{cgend}), 
without any extra terms like in (\ref{inter}-\ref{cg}). 
This is a nice confirmation 
of our HVBM results and underlines the consistency of the whole calculation. 
However, it needs to be emphasized that although our choice of the reading
points above seems the most reasonable one, other choices should be 
allowed. But if one insists on choosing the reading point
$-\gamma_5 \slsh{n}/4 n \cdot k$ also for graph (h) of the $C_F T_f$ 
part of $\Delta P_{qg}^{(1)}$ (rather than shifting it to the black
dot in Fig.~\ref{topolqg}), 
the result for this graph is no longer different from
the one in the HVBM scheme! Thus a genuinely different final answer
for $\Delta P_{qg}^{(1)}$ would be the consequence. We also found
that choosing a different reading point in the calculation of 
$\Delta C_q$ the result seems to change. It appears likely that these
problems indicate a certain incompatibility of the $\gamma_5$-scheme
of [\ref{KORN}] with the projection method we are using\footnote{On the
other hand, similar problems with the prescription of [\ref{KORN}] also
seem to have been the source of an error in a previous version of the 
OPE calculation of [\ref{MVN}].} in which, as
discussed above, traces sometimes are cut, with insertions of 
$\gamma_5$, by action of $\Proj$. The genuinely four-dimensional (i.e., 
non-anticommuting) $\gamma_5$ of [\ref{HVBM}] seems to be more 
appropriate here since it gives a unique answer without any extra
effort apart from the scheme transformation (\ref{strafo},\ref{zqq}).

We mention that we have also recalculated some graphs of the $C_F T_f$
part of $\Delta P_{qg}^{(1)}$, as well as the coefficient function 
$\Delta C_q$, using the prescription of [\ref{LV},\ref{NZ}],
i.e., eliminating $\gamma_5$ via Eq.~(\ref{larg5}) and contracting 
the Levi-Civita tensors in $d$ dimensions. This scheme is expected to 
be equivalent to the HVBM scheme [\ref{LV}]. In all cases we studied 
we indeed obtained the same answer as in the HVBM scheme without having
to take into account any $(d-4)$-dimensional scalar products which are
not present in this scheme since the metric tensors coming from the 
products of two $\epsilon$-tensors are taken to be $d$-dimensional.
This computational advantage is lost, however, because the trace operations
become considerably more involved due to the substitution 
(\ref{larg5}). Furthermore, it is not entirely clear how to generalize 
this scheme in order to deal with traces with two $\gamma_5$, i.e., how to 
establish equivalence to the HVBM prescription in this case. 
\subsection{Connection with dimensional reduction}
We finally discuss how our results in Eqs.~(\ref{p1ns}-\ref{cgend}) can 
be translated to dimensional reduction. As was explained in 
[\ref{SSK},\ref{KT}], the UV sectors of QCD in dimensional regularization 
and dimensional reduction are made to agree by introducing additional 
counterterms in the latter which include a finite renormalization of the 
strong charge. Once this is done, all remaining differences between
the results for a NLO quantity in dimensional regularization and in 
dimensional reduction can only be due to the effects of mass singularities.
They are fully accounted for [\ref{KT},\ref{KST},\ref{KAMAL}] by the 
differences between the $d$-dimensional LO splitting functions 
(as to be obtained in dimensional {\em regularization}) and the 
{\em four}-dimensional ones (corresponding to dimensional {\em reduction}).
This makes it very easy to transform our results in 
Eqs.~(\ref{p1ns}-\ref{cgend}) to dimensional reduction: We just need
to perform a factorization scheme transformation (\ref{strafo}), with
the $\Delta z_{ij}$ to be obtained from the parts $\sim \epsilon$ of the 
polarized $d$-dimensional LO splitting functions as obtained in the HVBM 
scheme. The latter read:
\beqn
\label{Pqqd}
\Delta P_{qq}^{(0),d=4-2 \epsilon}(x)&=&
C_F \Big\{\frac{2}{[1-x]_+}-1-x +3 \epsilon (1-x)
+\frac{3+\epsilon}{2}\delta(1-x)
\Big\} \\
\label{Pqgd}
\Delta P_{qg}^{(0),d=4-2\epsilon}(x)&=&2 T_f\left\{2x-1 
-2 \epsilon (1-x) \right\}  \\
\label{Pgqd}
\Delta P_{gq}^{(0),d=4-2 \epsilon}(x)&=&C_F \left\{ 2-x 
+2 \epsilon (1-x) \right\} \\
\label{Pggd}
\Delta P_{gg}^{(0),d=4-2 \epsilon}(x)&=&2 \nc 
\Big\{\frac{1}{[1-x]_+}-2 x+1 +2 \epsilon (1-x) \Big\}
+\left( \frac{\beta_0}{2} +\frac{N_C}{6} \epsilon\right) \delta(1-x)  \; .
\eeqn
However, care has to be taken: We have already performed the scheme 
transformation (\ref{zqq}) which effectively amounted to removing 
$4 \epsilon (1-x)$ from $\Delta P_{qq}^{(0),d=4-2 \epsilon}$, 
hereby rendering it equal to its unpolarized counterpart,
$P_{qq}^{(0),d=4-2 \epsilon}$ (see Eq.~(\ref{pqqdim})). We thus 
have to use 
\beq
\label{Pqqdd}
\Delta P_{qq}^{(0),d=4-2 \epsilon}(x)=
C_F \Big\{\frac{2}{[1-x]_+}-1-x - \epsilon (1-x)
+\frac{3+\epsilon}{2}\delta(1-x)
\Big\} 
\eeq
instead of (\ref{Pqqd}). From the parts $\sim \epsilon$ in 
Eqs.~(\ref{Pqgd}-\ref{Pqqdd}) we can now read off the $\Delta z_{ij}$ 
to be used in the new scheme transformation (\ref{strafo}):
\beqn
&&\Delta z_{qq} (x) = \frac{C_F}{4} \left( (1-x) - \frac{1}{2} \delta (1-x) 
\right) \;\; , \;\;\;  
\Delta z_{qg} (x) = T_f (1-x)  \; ,  \nonumber \\
&&\Delta z_{gq} (x) = -\frac{C_F}{2} (1-x) \;\; , \;\;\;
\Delta z_{gg} (x) = -N_C \left( (1-x) + \frac{1}{24} \delta (1-x) \right)
\; . 
\eeqn
Inserting these into (\ref{strafo}), one finds:
\beqn
\label{drqq}
\Delta P_{qq}^{(1),{\rm DR}} - \Delta P_{qq}^{(1)} (x) &=&
-\frac{\beta_0 C_F}{2} \left( (1-x) - \frac{1}{2} \delta (1-x) \right) 
-4 C_F T_f (1-x) \ln x  \; , \\
\label{drqg}
\Delta P_{qg}^{(1),{\rm DR}} - \Delta P_{qg}^{(1)} (x) &=&
C_F T_f \left( 9-10x +6 \ln x -8 (1-x) \ln (1-x) \right) \nonumber \\
&+& N_C T_f \left( -\frac{25}{3} +\frac{26}{3} x +8 (1-x) \left( 
\ln (1-x) - \ln x \right) \right) \; , \\
\label{drgq}
\Delta P_{gq}^{(1),{\rm DR}} - \Delta P_{gq}^{(1)} (x) &=&
C_F^2 \left( 5-\frac{9}{2} x -4 (1-x) \ln (1-x) +(4-x) \ln x  \right) 
\nonumber \\
&+& C_F N_C \left( 3-\frac{19}{6} x +4 (1-x) \ln (1-x) \right) - 
\frac{8}{3} C_F T_f (1-x) \; , \\
\label{drgg}
\Delta P_{gg}^{(1),{\rm DR}} - \Delta P_{gg}^{(1)} (x) &=&
2 \beta_0 N_C \left( (1-x) +\frac{1}{24} \delta (1-x) \right) 
+4 C_F T_f (1-x) \ln x \; ,
\eeqn
where the $\Delta P_{ij}^{(1)}$ are our final results in 
Eqs.~(\ref{p1ns}-\ref{p1gg}) and the $\Delta P_{ij}^{(1),{\rm DR}}$
are the polarized NLO ($\overline{\mbox{MS}}$)
splitting functions in dimensional reduction.
For the $qq$ sector we have only written down the entry of the 
singlet matrix, $\Delta P_{qq}^{(1)}=\Delta P_{qq}^{+,(1)}+
\Delta P_{qq}^{S,(1)}$ (see Eq.~(\ref{qqs})). Furthermore, we 
have not written down the short-distance cross sections 
since their transformation is trivial.
We mention however, that the effect of $\Delta z_{qg}$ on $\Delta C_g$
is to remove the term $+2 (1-x)$ at the end of (\ref{cgend}),
which corresponds to a factorization scheme in which gluons contribute 
to the first moment ($x$-integral) of the structure function $g_1$. 
Such a factorization scheme was suggested in [\ref{ar}]. Since 
furthermore the first moments of $\Delta P_{qq}^{(1),{\rm DR}}$
and $\Delta P_{qg}^{(1),{\rm DR}}$ vanish, it follows that 
the total polarization of quarks and antiquarks, $\int_0^1 dx
\Delta \Sigma^{\rm DR} (x,Q^2)$, is $Q^2$-independent when defined
in dimensional reduction.

As a little cross-check on our findings in Eqs.~(\ref{drqq}-\ref{drgg})
we have recalculated all polarized NLO splitting functions, using
all matrix elements in {\em four} dimensions (i.e., setting 
$\epsilon$ and $(d-4)$-dimensional scalar products to zero), 
but performing loop and 
phase space integrations in $d$ dimensions {\em after} all spin algebra 
has been done. Subtracting the corresponding results from our HVBM ones, 
Eqs.~(\ref{p1ns}-\ref{p1gg}), we reproduced all logarithmic terms in
Eqs.~(\ref{drqq}-\ref{drgg}), leaving only some non-logarithmic pieces
(and of course the $\delta$-functions) unaccounted for. It is precisely 
such non-logarithmic terms one expects to be generated by inclusion 
of the additional finite UV-counterterms needed for a full and proper 
calculation in dimensional reduction [\ref{SSK},\ref{KT}].   

The main reason for performing our exercise concerning dimensional 
reduction is that the results for $\Delta P_{ij}^{(1),{\rm DR}}$ to be
obtained from Eqs.~(\ref{drqq}-\ref{drgg}),(\ref{p1ns}-\ref{p1gg})
fulfil a remarkable relation if one sets $C_F=N_C=2 T_f\equiv N$
(see also [\ref{MVN}]):
\beq
\Delta P_{qq}^{(1),{\rm DR}} (x) + \Delta P_{gq}^{(1),{\rm DR}} (x) -
\Delta P_{qg}^{(1),{\rm DR}} (x) - \Delta P_{gg}^{(1),{\rm DR}} (x) \equiv 0
\eeq
which is known to hold for LO splitting functions and also for the 
unpolarized NLO splitting functions in dimensional reduction 
[\ref{fa},\ref{KST}], and is expected from supersymmetry.
\section{Summary}
We have presented a detailed description of our calculation 
[\ref{Vogelsang}] of the spin-dependent NLO GLAP splitting functions in 
the method of [\ref{EGMPR},\ref{CFP},\ref{FP}]. The main calculation has 
been performed using the $\gamma_5$-prescription of [\ref{HVBM}], but we 
have also discussed the results one obtains when using a fully anticommuting 
$\gamma_5$, giving up cyclicity of the Dirac-trace [\ref{KORN}].
The results for the two schemes turn out to be the same for a reasonably 
chosen reading point in the latter scheme, but the HVBM scheme 
appears to be the safer and the more straightforward prescription.
Our final result confirms the previous OPE results of [\ref{MVN}]. We 
have also provided the connection with dimensional reduction, in which 
the NLO splitting functions satisfy a simple supersymmetric relation.
\section*{Acknowledgements}
I am thankful to R.K.~Ellis and M.~Stratmann for helpful discussions.
\begin{figure}[htb]
\vspace{16cm}
\includegraphics{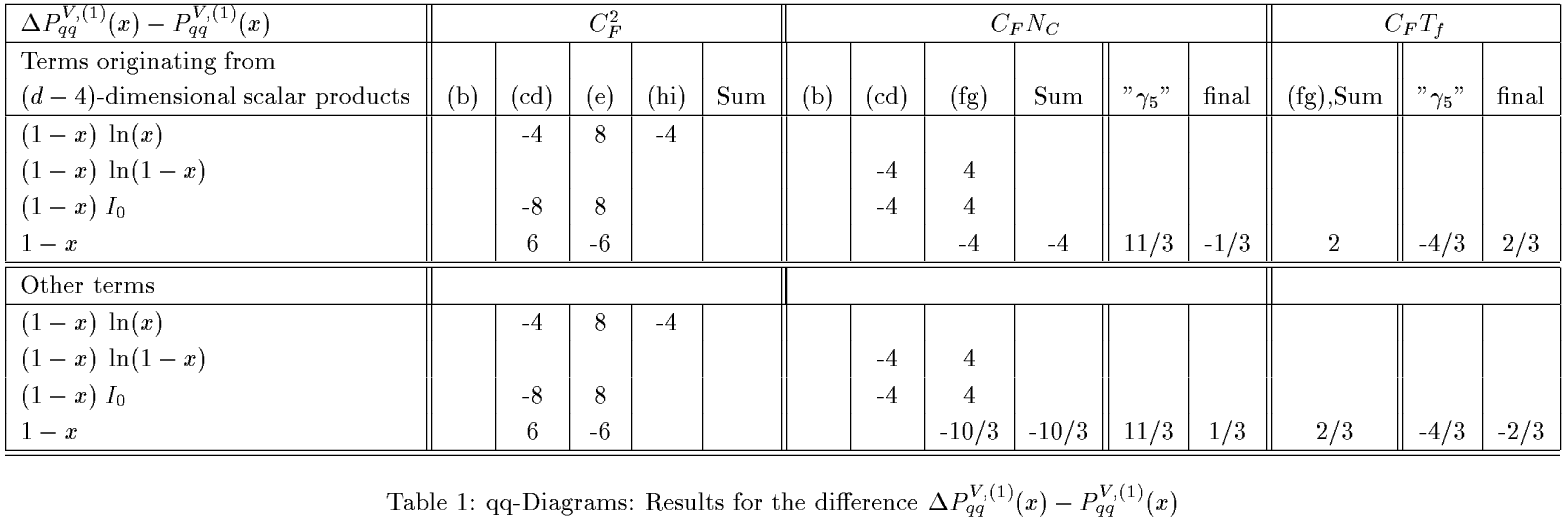}
\end{figure}
\begin{figure}[htb]
\vspace{16cm}
\includegraphics{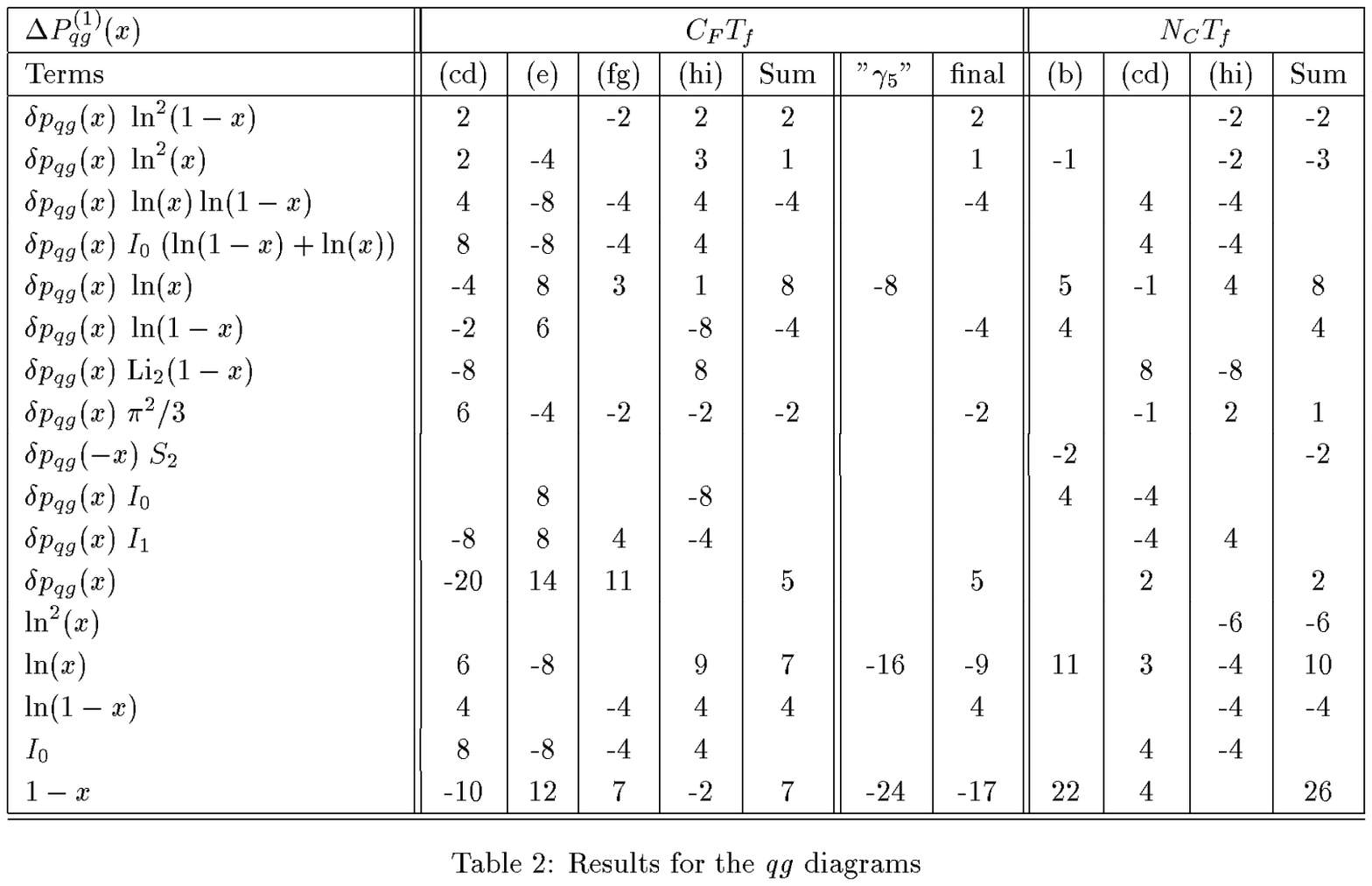}
\end{figure}
\begin{figure}[htb]
\vspace{16cm}
\includegraphics{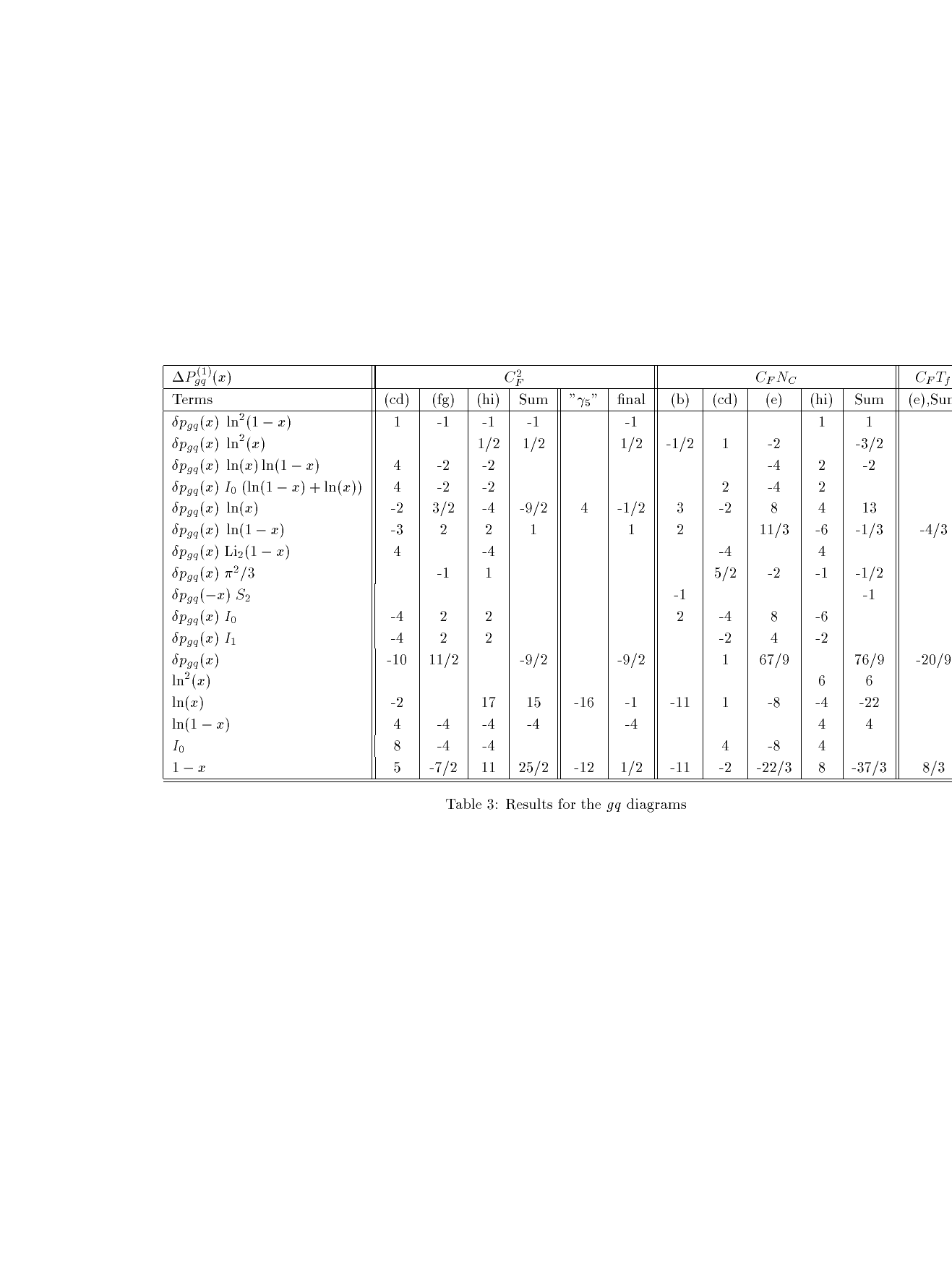}
\end{figure}
\begin{figure}[htb]
\vspace{16cm}
\includegraphics{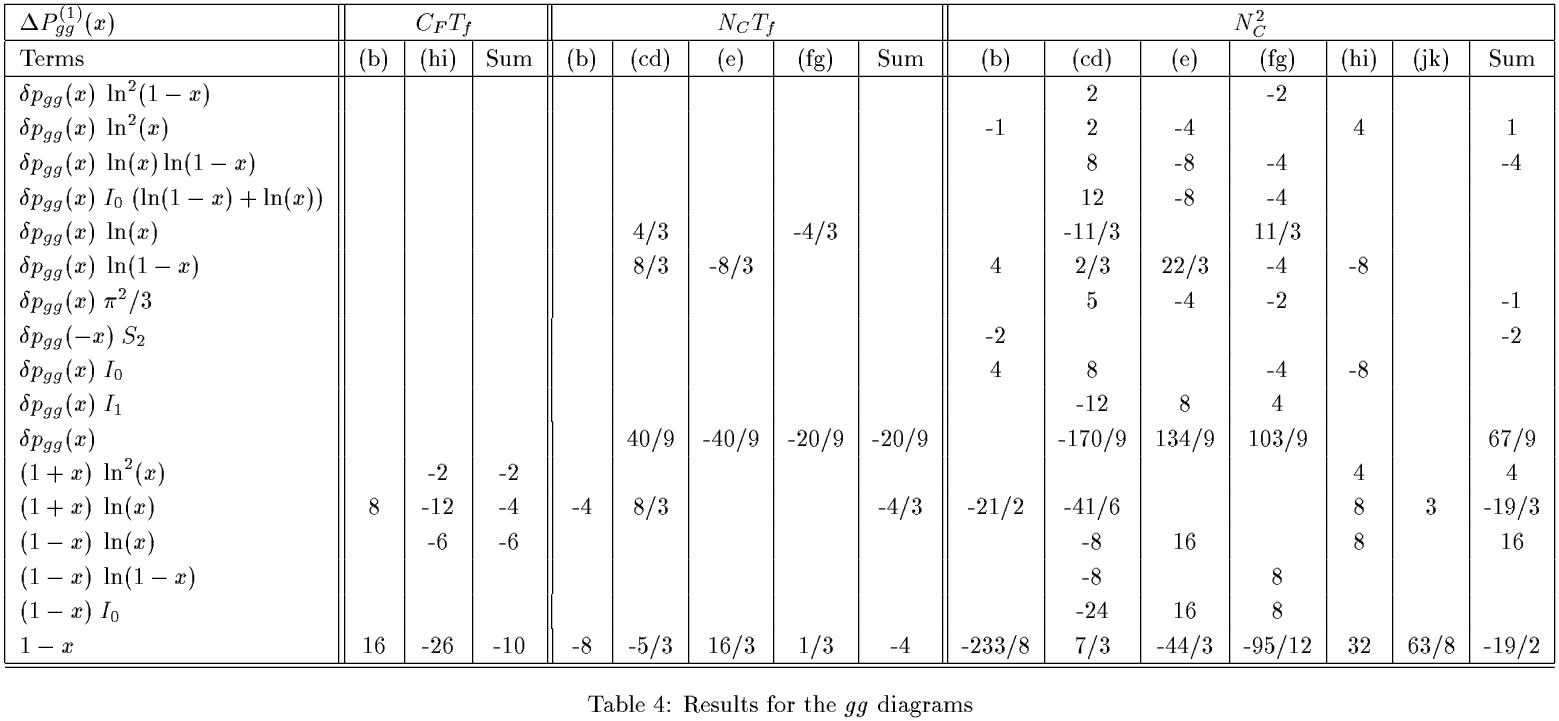}
\end{figure}
\setcounter{table}{6}
\clearpage
\section*{Appendix}
\begin{figure}[htb]
\vspace{5cm}
\includegraphics{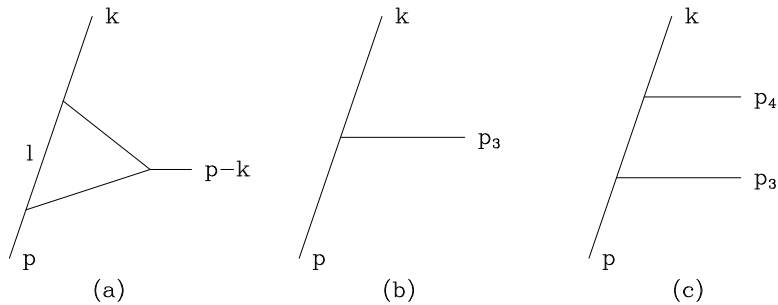}
\caption{(a) Vertex correction graph (b) One parton emission (c) 
Two parton emission.}
\label{pd}
\end{figure}

\appendix
\setcounter{equation}{0}
\renewcommand{\theequation}{A.\arabic{equation}}
Almost all technical ingredients we need, like two- and three-point
functions and three-particle phase space integrals, have recently 
been given in the documentation [\ref{EV}] of the unpolarized
calculation. We only need to consider some new details related to the
treatment of $\gamma_5$ and $\epsilon^{\mu\nu\rho\sigma}$ in 
$d$ dimensions, which can introduce explicit dependence of the 
matrix element squared on scalar products of the nonphysical
($d-4$-dimensional) components of vectors (see section 2.4). 
It is most convenient to 
work in the IMF parametrization of the momenta [\ref{CFP}]:
\begin{eqnarray}
p &=& (\:\: P\:\:,\:\:\vec{0}_{xy}\:\: ,\:\: P\:\: ,\:\:
\vec{0}_{d-4}\:\: ) \:\:\: , \nonumber \\
n &=& (\:\: \frac{pn}{2P}\:\: , \:\: \vec{0}_{xy} \:\: ,
\:\: -\frac{pn}{2P} \:\: ,\:\: \vec{0}_{d-4} \:\: ) 
\:\:\: , \nonumber \\
k &=& \left( \:\: xP + \frac{k^2+\tilde{k}^2}{4xP} \:\: , 
\:\: \vec{k}_T \:\: ,
\:\: xP - \frac{k^2+\tilde{k}^2}{4xP}, \:\: 
\hat{\vec{k}} \:\: \right) \:\:\: , \label{a1}
\end{eqnarray}
where $p$ represents the incoming and $k$ the outgoing ('observed') parton,
with 
\beq
\tilde{k}^2 \equiv  k_x^2+k_y^2+(\hat{\vec{k}})^2 \equiv k_T^2+\hat{k}^2  
\eeq 
being the total transverse momentum squared of $k$ relative to  
the longitudinal axis defined by $p,n$. Here we have explicitly 
introduced the ($d-4$)-dimensional components of $k$ with
\beq
\hat{k}^2 \equiv (\hat{\vec{k}})^2 = -\hat{g}_{\mu\nu} k^{\mu}k^{\nu}  
 \; , 
\eeq
$\hat{g}_{\mu\nu}$ being the $(d-4)$-dimensional metric tensor of 
section 2.4. The labelling of the other momenta is fixed in Fig.~\ref{pd}.
\section{Virtual Integrals}
\renewcommand{\theequation}{\thesection.\arabic{equation}}
The virtual diagrams always comprise emission of only one real massless
particle (the rung of the ladder) from the process $p\rightarrow k+p_3$ 
(see Fig.~\ref{pd}(a),(b)). The corresponding phase space reads (including 
the integration over the 'observed' parton $k$):
\beqn
PS^{(1)} &=&
\int d^d k \; x \delta (x- \frac{n \cdot k}{pn})
\int \frac{d^{d}p_3}{(2 \pi)^{d-1}} \delta^+(p_3^2)  \; 
(2\pi)^d \delta^d(p-p_3-k) 
\nonumber \\  \label{ps1} 
&=& \frac{x \pi^{2-\epsilon}}{\Gamma (1-\epsilon)} (1-x)^{-\epsilon}
\int_0^{Q^2} d|k^2| |k^2|^{-\epsilon} \Bigg(
(-\epsilon) \int_0^1 d\hat{\kappa} \; \hat{\kappa}^{-1-\epsilon} \Bigg) \; ,
\eeqn
where $\hat{\kappa}$ is defined by
\beq
\hat{k}^2 = |k^2| (1-x) \hat{\kappa} \; .
\eeq
The last integral in Eq.~(\ref{ps1}) has been written in such a way that
it is unity if there is no dependence on $\hat{\kappa}$, i.e., $\hat{k}^2$.
As we can also see, any term proportional $\hat{k}^2$ in the matrix 
element will give a factor $\epsilon$ after performing the 
$\hat{\kappa}$-integration. 

When performing a loop calculation in the polarized case, the matrix 
element will in general not only depend on $\hat{k}^2$, but also on scalar
products involving the $(d-4)$-dimensional components of the loop 
momentum $l$, like $\hat{l}^2$, $\hat{\vec{k}}\cdot \hat{\vec{l}}$.
One can always write, e.g., $\hat{\vec{l}}\cdot \hat{\vec{k}}$ as
$-\hat{g}_{\mu\nu} l^{\mu} k^{\nu}$, then perform the loop 
integration, and contract the result with $-\hat{g}_{\mu\nu} k^{\nu}$ 
afterwards. Obviously, this requires knowledge of vectorial two- or 
three-point functions (and tensorial ones when $\hat{l}^2=-\hat{g}_{\mu\nu}
l^{\mu}l^{\nu}$ appears). These can be easily obtained from the
scalar ones presented in [\ref{EV}]. The only new feature is that,
unlike in the unpolarized calculation, we also need the vectorial 
three-point functions with a light-cone gauge propagator,
\beq
J_{3,n}^{\mu} \equiv 
\int \frac{d^d l}{(2 \pi)^d} \mbox{PV}\Bigg( \frac{p_+}{\lp}\Bigg)
\frac{l^{\mu}} {(l^2+i \varepsilon)((l-k)^2+i \varepsilon)
((l-p)^2+i \varepsilon)}  \; ,
\eeq
where $p_+ = pn$, $l_+ = n\cdot l$, and 'PV' denotes use 
of the principal value prescription (\ref{PPprescription}) in order
to regularize the singularities of the light-cone gauge propagator. 
More precisely, since $J_{3,n}^{\mu}$ will be contracted with 
$-\hat{g}_{\mu\nu} k^{\nu}$ afterwards, we only need its part $\sim
k^{\mu}$, as $p$, $n$ do not possess non-vanishing components in their 
$(d-4)$-dimensional parts.
Assuming Lorentz covariance of the integral one finds
\beq \label{vector}
J_{3,n}^{\mu} = \frac{1}{2 (1-x)} \left( J_{3,n} - J_3
+ \frac{1}{|k^2|} J_{2,n}  \right)  k^{\mu} + \ldots  \; ,
\eeq
where $J_{2,n}$, $J_{3,n}$ are the scalar two- and three-point functions
with a light-cone gauge propagator as given in Eqs.~(A.4) and (A.14) of
[\ref{EV}], respectively, 
and $J_3$ is the ordinary scalar three-point function 
of (A.13) in [\ref{EV}]. The dots in (\ref{vector}) indicate 
the contributions proportional to $p^{\mu}$ and $n^{\mu}$, which we 
do not need. We finally note that in our previous paper [\ref{Vogelsang}]
on the polarized NLO splitting functions we were able to determine
the results for most of the virtual graphs just from their contributions
in the {\em un}polarized case. The results in Appendix A of 
[\ref{EV}] and the ones presented here now enable a straightforward direct
calculation of all virtual contributions also for the polarized case.
\section{Real Integrals}
The real emission processes $p \rightarrow k+p_3+p_4$ (see Fig.~\ref{pd}(c))
comprise integration over the momenta of two unobserved real massless
particles. In addition to (\ref{a1}) we define 
\beq
p_3 = (\:\: p_3^0 \:\: ,\:\: \vec{p}_3^{\; xy} \:\: ,
\:\: p_3^z \:\: , \:\: \hat{\vec{p}}_3 \:\: ) \:\:\: , \label{p3def}
\eeq
introducing the $(d-4)$-dimensional components of $p_3$ which we 
split into a part $\hat{p}_3^{\: \parallel}$ parallel to those of $k$ 
and a transverse part $\hat{p}_3^{\perp}$. The phase space then
reads (again including the integration over the 'observed' parton $k$):
\beqn
PS^{(2)} &=& \int d^d k x \delta \left (x-\frac{n\cdot k}{pn} \right)
\int \frac{d^d p_3}{(2\pi)^{d-1}} \delta^+ (p_3^2)
\int \frac{d^d p_4}{(2\pi)^{d-1}} \delta^+ (p_4^2) 
(2\pi)^d \delta^d (p-k-p_3-p_4) \nonumber \\
&=& \frac{1}{2^{4+2 \epsilon} \pi \Gamma (1-2 \epsilon)}
x^{\epsilon} (1-x)^{1-2 \epsilon} \int_0^{Q^2} d|k^2| 
|k^2|^{1-2 \epsilon} \label{ps2} \\
&&\times \int_0^1 d\tilde{\kappa} 
\left( \tilde{\kappa} (1-\tilde{\kappa}) \right)^{-\epsilon} \int_0^1
dw \left( w (1-w) \right)^{-\epsilon} \int_0^1 dv \left( v (1-v)  
\right)^{-\frac{1}{2}-\epsilon} \nonumber \\
&&\times \Bigg(  (-\epsilon) \int_0^1 d\hat{\kappa}
\hat{\kappa}^{-1-\epsilon} \Bigg) \Bigg( (-\frac{1}{2}-\epsilon)
\int_0^1 d\lambda^{\perp} \left( \lambda^{\perp}\right)^{-3/2-\epsilon}
\Bigg) \Bigg( \frac{1}{\pi} \int_0^1 d\lambda^{\parallel} 
\left( \lambda^{\parallel} (1-\lambda^{\parallel}) \right)^{-1/2} \Bigg)
\nonumber 
\eeqn
where we have defined
\begin{eqnarray}
\hat{k}^2 &=& |k^2| (1-x) \hat{\kappa} \tilde{\kappa} 
\:\:\: , \nonumber \\ 
\tilde{k}^2 &=& |k^2| (1-x) \tilde{\kappa} \:\:\: , \nonumber \\ 
p_3^0 + p_3^z &=& 2 P (1-x) w 
=2 P \frac{n\cdot p_3}{pn} \:\:\: , \nonumber \\ 
(p_3^0)^2-(p_3^z)^2 &=& c_1^2 + v (c_2^2-c_1^2) 
= \frac{1}{P} (p_3^0+p_3^z) (p \cdot p_3) \:\:\: , \nonumber \\ 
\hat{p}_3^{\: \parallel} &=& \lambda_1 + \lambda^{\parallel} 
(\lambda_2 - \lambda_1) \:\:\: , \nonumber \\         
(\hat{p}_3^{\perp})^2 &=& v (1-v) \left( c_1+c_2 \right)^2 
\lambda^{\perp} 
\end{eqnarray}
with 
\begin{eqnarray}
c_{1,2} &\equiv& \sqrt{\frac{|k^2| (1-x) w}{x}} \Bigg[
\sqrt{(1-w) (1-\tilde{\kappa})}\mp \sqrt{x w \tilde{\kappa}} \Bigg]
\:\:\: , \nonumber \\
\lambda_{1,2} &=& -\frac{1}{2} \frac{\hat{\kappa}}{\hat{k}w}
\left( (p_3^0)^2-(p_3^z)^2 -c_1 c_2 \right) 
\mp (c_1 + c_2 )\sqrt{(1-\hat{\kappa})(1-\lambda^{\perp})v (1-v)}
\:\:\: .         
\end{eqnarray}
Again the last three integrals in Eq.~(\ref{ps2}) have been written in such 
a way that they are all unity if there is no dependence on 
($d-4$)-dimensional scalar products. If present, such terms only give 
contributions proportional to $\epsilon$ after the last three integrals 
have been performed. The advantage of writing $PS^{(2)}$ in this way
and performing the three last integrals first is that the result
of these integrations can always be expressed in terms of usual
($d$-dimensional) scalar products. For instance, 
\beqn  \label{express}
1 &\longrightarrow& 1 \; , \nonumber \\
\hat{k}^2 &\longrightarrow& -\frac{\epsilon}{1-\epsilon} \left( 
|k^2| +2 x (p\cdot k) \right) \; , \nonumber \\
\hat{p}_3^2 \equiv ( \hat{p}_3^{\: \parallel} )^2 + 
\left( \hat{p}_3^{\perp} \right)^2
&\longrightarrow& -\frac{2 \epsilon}{1-\epsilon}
(p\cdot p_3) \frac{n\cdot p_3}{pn} \; , \\
-\hat{g}_{\mu\nu} k^{\mu} p_3^{\nu} \equiv \hat{k} \hat{p}_3^{\:
\parallel} &\longrightarrow&
-\frac{\epsilon}{1-\epsilon} \left( -(1-x) (p\cdot p_3) - \frac{|k^2|}{2}
-(p\cdot k) +(p \cdot k) \frac{n \cdot p_3}{pn} \right) \nonumber 
\eeqn
after integration over $\hat{\kappa}$, $\lambda^{\perp}$ and 
$\lambda^{\parallel}$. The other two 
$(d-4)$-dimensional scalar products that appear in the calculation,
$\hat{k}^2 \hat{p}_3^2$ and $(\hat{k} \hat{p}_3^{\: \parallel})^2$,
have lengthy expressions which can be straightforwardly 
obtained from (\ref{ps2}). Thus, after performing the last three 
integrations in (\ref{ps2}), there are only terms left in the matrix 
element which are familiar from the unpolarized case. The matrix element can
then be further integrated using the results of Appendix B of
[\ref{EV}] or, equivalently, doing the remaining integrals 
in Eq.~(\ref{ps2}).
\begin{reflist}
\item \label{AlPa}
G. Altarelli and G. Parisi, \np{B126}{298}{77}.
\item \label{GLAP}
Yu.~L.~Dokshitzer \JETP{46}{641}{77}; \newline
\cf~L.~N.~ Lipatov, \sj{20}{95}{75}; \newline
V.N.~Gribov and L.N.~ Lipatov, \sj{15}{438}{72}.
\item \label{FLOR} E.G. Floratos, D.A. Ross, and C.T. Sachrajda,
\np{B129}{66}{77}; E: {\bf B139} (1978) 545; \np{B152}{493}{79}; \\
see also:
A. Gonzales-Arroyo, C. Lopez, and F.J. Yndurain, \np{B153}{161}{79};
A. Gonzales-Arroyo and C. Lopez, \np{B166}{429}{80};
E.G. Floratos, C. Kounnas, and R. Lacaze, \pl{98B}{89,285}{81};
\np{B192}{417}{81}.
\item \label{CFP}
G. Curci, W. Furmanski and R. Petronzio, \np{B175}{27}{80}.
\item \label{FP}
W. Furmanski and R. Petronzio, \pl{97B}{437}{80}.
\item \label{EGMPR}
R.K. Ellis, H. Georgi, M. Machacek, H.D. Politzer, and G.G. Ross, 
\pl{78B}{281}{78}; \np{B152}{285}{79}.
\item \label{HamVN}
R. Hamberg, PhD Thesis, University of Leiden, 1991; \newline
R. Hamberg and W. van Neerven, \np{B379}{143}{92}.
\item \label{EV}
R.K. Ellis and W. Vogelsang, CERN-TH/96--50, RAL-TR-96-012, hep-ph/9602356.
\item \label{Leib}
G. Leibbrandt, \pr{D29}{1699}{84}; \\
For an overview see also: G. Leibbrandt, \rmp{59}{1067}{87}.
\item \label{AT}
A. Andrasi and J. C. Taylor, \np{B310}{222}{88}.
\item \label{Vogelsang}
W. Vogelsang, Rutherford report, RAL-TR-95-071, hep-ph/9512218.
\item \label{MVN} 
R.\ Mertig and W.L.\ van Neerven, Univ.\ Leiden
INLO-PUB-6/95 and NIKHEF-H/95-031, June 1995, November 1995 (revised).
\item \label{HVBM} 
G. 't Hooft and M. Veltman, \np{B44}{189}{72};
P. Breitenlohner and D. Maison, {\it Comm. Math. Phys.} {\bf 52} (1977) 11.
\item \label{KORN} 
J.G. K\"{o}rner, D. Kreimer, and K. Schilcher, \zp{C54}{503}{92}.
\item \label{LV}
S.A. Larin and J.A.M. Vermaseren, \pl{B259}{345}{91};
S.A. Larin, \pl{B303}{113}{93}.
\item \label{CFH} 
M. Chanowitz, M. Furman, and I. Hinchliffe, 
\np{B159}{225}{79}.
\item \label{TRAC} 
M. Jamin and M.E. Lautenbacher, 
TU M\"{u}nchen report TUM-T31-20/91.
\item \label{NZ} 
E.B. Zijlstra and W.L. van Neerven, \pl{B297}{377}{92};
\np{B417}{61}{94}; E: {\bf B426} (1994) 245.
\item \label{siegel}
W. Siegel, \pl{84B}{193}{79}.
\item \label{SSK} 
G.A. Schuler, S. Sakakibara, and J.G. K\"{o}rner, \pl{B194}{125}{87}.
\item \label{KT} 
J.G. K\"{o}rner and M.M. Tung, \zp{C64}{255}{94}.
\item \label{KST} 
Z. Kunszt, A. Signer, and Z. Trocsanyi, 
\np{B411}{397}{94}.
\item \label{KAMAL} 
B. Kamal, \pr{D53}{1142}{96}. 
\item \label{AhRo} 
M.A. Ahmed and G.G. Ross, \np{B111}{441}{76}.
\item \label{Wada} 
R.T. Herrod and S. Wada, \pl{96B}{195}{80}.
\item \label{dd} 
A. Devoto and D.W. Duke, {\it Nuov. Cim. (Riv.)} {\bf 7} 
(1984) 1.
\item \label{grsv} 
M.\ Gl\"{u}ck, E.\ Reya, M. Stratmann, and W.\ Vogelsang, 
Univ.\ Dortmund DO-TH 95/13, Rutherford RAL-TR-95-042, to appear in
{\it Phys. Rev.} {\bf D}.
\item \label{ms} 
M. Stratmann, W. Vogelsang, and A. Weber, \pr{D53}{138}{96}.
\item \label{kod} 
J. Kodaira, S. Matsuda, K. Sasaki, and T. Uematsu,
\np{B159}{99}{79}.
\item \label{GR} 
M. Gl\"{u}ck and E. Reya, \pr{D25}{1211}{82}.
\item \label{Alex} 
A. Weber, \np{B382}{63}{92}.
\item \label{ar}
G. Altarelli and G.G. Ross, \pl{B212}{391}{88};
G. Altarelli and W.J. Stirling, {\it Particle World} {\bf 1} (1989) 40;
G. Altarelli and B. Lampe, \zp{C47}{315}{90}. 
\item \label{fa} 
I. Antoniadis and E.G. Floratos, \np{B191}{217}{81}.
\end{reflist}
\end{document}